# Graphene-Based Spintronic Components

Minggang Zeng,†,‡ Lei Shen,*,† Haibin Su,¶,§ Miao Zhou,† Chun Zhang,†,‖ and Yuanping Feng*,†

*Department of Physics, 2 Science Drive 3, National University of Singapore, Singapore 117542, Singapore, NanoCore, 5A Engineering Drive 4, National University of Singapore, Singapore 117576, Singapore, Division of Materials Science, Nanyang Technological University, 50 Nanyang Avenue, Singapore 639798, Singapore, Institute of High Performance Computing, 1 Fusionopolis Way, Connexis 138632, Singapore, and Department of Chemistry, 3 Science Drive 3, National University of Singapore, Singapore 117543, Singapore*

E-mail: shenlei@nus.edu.sg ; phyfyp@nus.edu.sg

## Abstract

A major challenge of spintronics is in generating, controlling and detecting spin-polarized current. Manipulation of spin-polarized current, in particular, is difficult. We demonstrate here, based on calculated transport properties of graphene nanoribbons, that nearly $\pm 100$ % spin-polarized current can be generated in zigzag graphene nanoribbons (ZGNRs) and tuned by a source-drain voltage in the bipolar spin diode, in addition to magnetic configurations of the electrodes. This unusual transport property is attributed to the intrinsic transmission selection rule of the spin subbands near the Fermi level in ZGNRs. The simultaneous control of spin current by the bias voltage and the magnetic configurations of the electrodes provides an opportunity to implement a whole range of spintronics devices. We propose theoretical designs for a complete set of basic spintronic devices, including bipolar spin diode, transistor and logic gates, based on ZGNRs.

## Graphical TOC

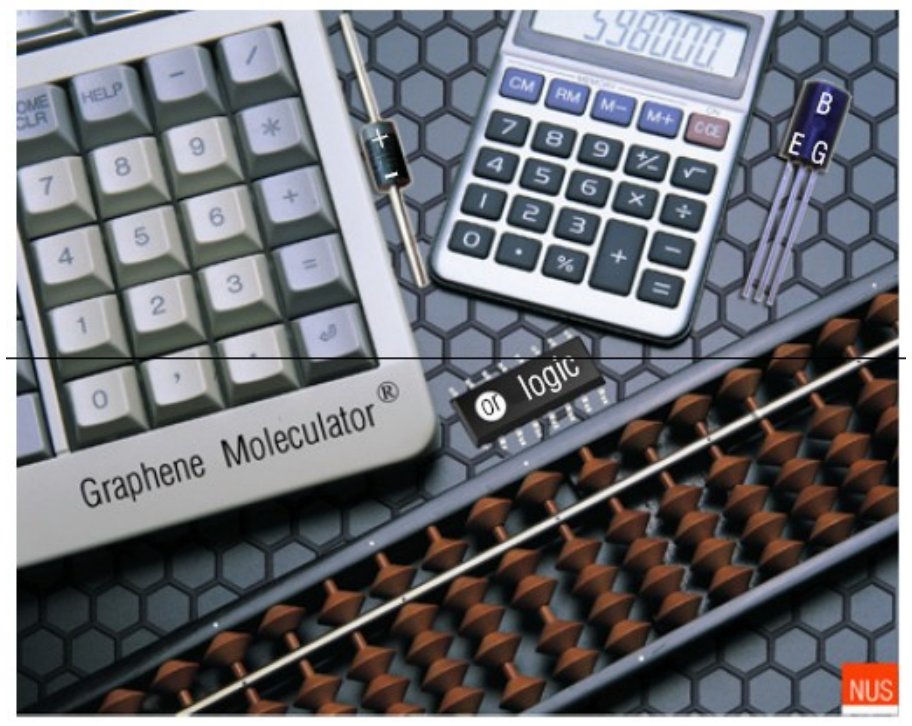

## Introduction

Spintronics, a new type of electronics that seeks to exploit the spin degree of freedom of an electron in addition to its charge, offers one of the most promising solutions for future high operating speed and energy-saving electronic devices.[1] The major challenge of spintronics is the difficulty in generating, controlling and detecting spin-polarized current. This situation is expected to change with successful development of spintronics devices based on graphene, an atomically thin carbon sheet.[2,3] Since graphene was discovered experimentally in 2004,[2] its applications in future electronics and spintronics have been a major research focus and to date, a number of major break-throughs have been made. Recently, spin injection in graphene has been demonstrated using cobalt electrode at room temperature.[4] More recently, Han *et al.* proposed that the efficiency of spin injection from a ferromagnetic electrode into graphene can be enhanced by a tunnel barrier.[5] Spin detection in graphene has also been realized using a non-local measurement method in which spin-up and spin-down currents can be distinguished and measured independently.[4] Moreover, room-temperature ferromagnetism of graphene was demonstrated byWang *et al:*.[6] Graphene nanoribbon is also known to have long spin diffusion length, spin relaxation time and electron spin coherence time.[7,8] All of these show that graphene hold the promise of spintronics.[9,10] A variety of graphene-based spintronics devices have been proposed.[11–16] For example, A graphene nanoribbon-based giant magnetoresistance (GMR) device has been proposed theoretically and realized experimentally which is useful for information storage.[11,12] Wang *et al.* also experimentally observed large MR in graphene-based devices.[13] However, most of these studies have been on individual component and focused mainly on devices for information storage. There has not been any implementation for the full range of spintronic devices including logic gates. Developments of multi-functional graphene-based spintronics components, such as devices that offer effective manipulation of spin-polarized current, are crucial for a complete realization of spintronics.

In this Letter, we explore possible control of spin-polarized current and logic operations in zigzag graphene nanoribbons (ZGNRs) based devices. We demonstrate that the spin polarization of the current in a ZGNR-based two-terminal device can be tuned by a source-drain voltage which is due to a symmetry selection rule, and the device functions as a bipolar spin diode. Spin polarized current is normally produced using "half-metallic" materials, such as, chemical functional groups decorated tri-wing graphene nanoribbons,[17] silicon-based half-metallic metal-encapsulated nanotube[18] and charge-injected single-layer hexagonal group III/V semiconductors.[19] Generation of spin polarized current using intrinsic symmetry selection rule proposed here is a totally different concept. For the first time, we also propose the use of magnetic configurations [$\pm$1,0], *i.e.,* one electrode is magnetized while the other is non-magnetic in the design, in addition to the conventional parallel[1,1]/anti-parallel[1,-1] magnetic configurations of GMR devices. The biggest advantage of the [$\pm$1,0] magnetic configurations is that it allows reduction of design complexity since only one electrode needs to be magnetized. It also avoids direct magnetic interaction between the two magnetic electrodes if the central part of nano-devices is very small. These novelties of graphene-based spintronic components offer flexibility in manipulating spin current and allow the designs of spintronic devices which provide the essential components for future spintronic integrated circuit.

# Results and discussion

## Model and methodology

Spin-dependent quantum transport calculations were carried out within the framework of DFT[20] and non-equilibrium Green's function method which has been widely used in studies of electron transport.[21] The computational details are given in the Supporting Information. Figures 1(a) and 1(b) show schematic configurations of a two-terminal ZGNR-based spin device. Magnetization of the ZGNR electrodes can be controlled by various methods, such as an external magnetic field[11] or an electrified solenoid and can be set to 1 (magnetization along the $+y$ direction shown in Fig. 1(a)), 0 (non-magnetization), or −1 (magnetization along $-y$ direction). We denote the magnetization of the left (right) electrode by $M_L$ ($M_R$) (=1; 0; or −1). Current flows along $z$ ($-z$) direction is defined as the + (−) direction, corresponding to a positive (negative) applied bias voltage. A ZGNR with $N$ zigzag chains is denoted by $N$-ZGNR.[22] In this Letter, we focus on the case in which $N$ is even since a previous study has concluded that only ZGNR with an even number of zigzag chain shows the transmission selection rule, which is related to the symmetry of ZGNRs.[23] Moreover, the width of the ZGNR will not affect our main conclusions [see Supporting Information Figure 1]. Thus, the ZGNR with 8 zigzag chains is used in our calculations.

## Bipolar spin diode

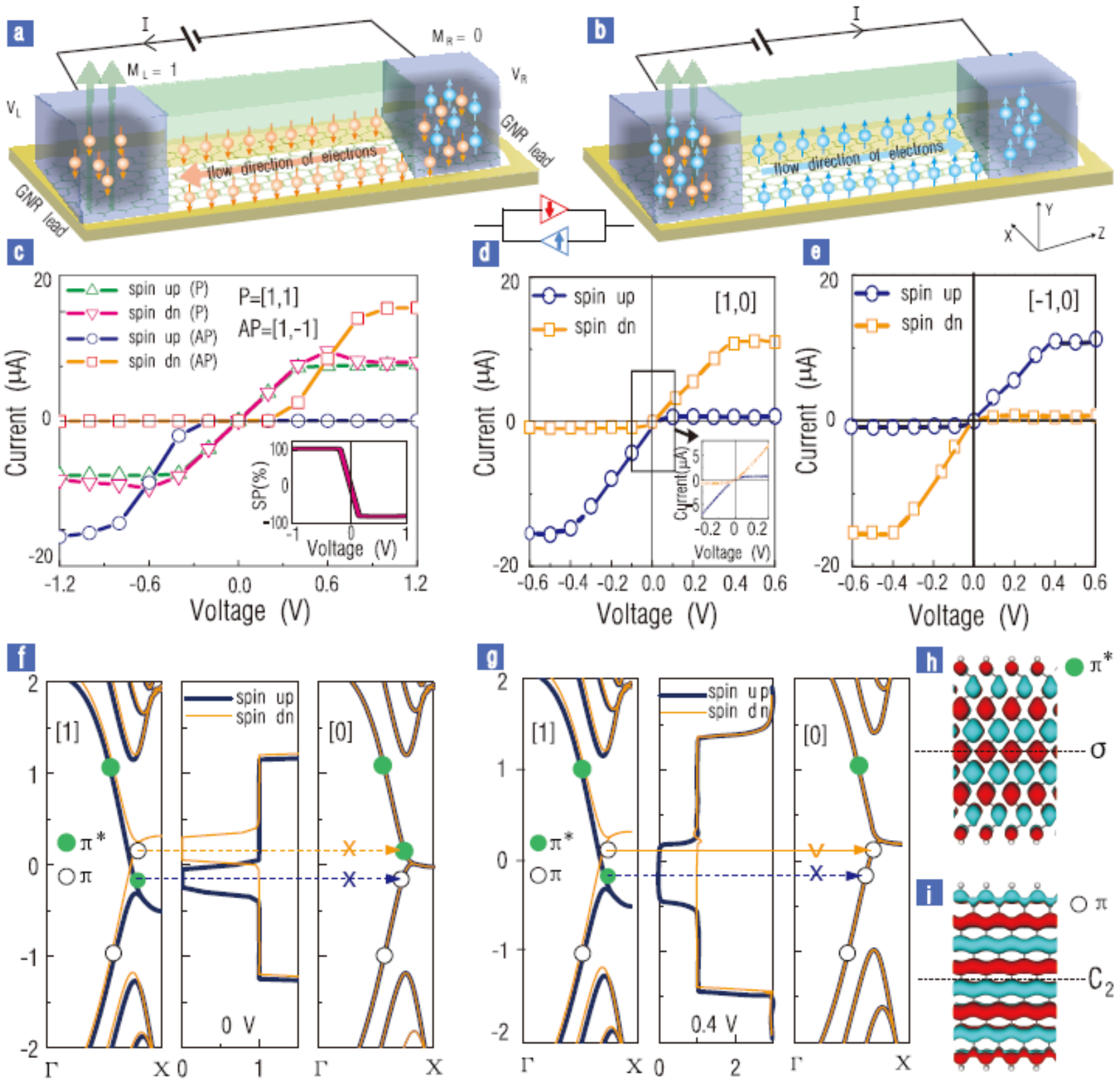

Figure 1: (Color online) Schematic diagram of ZGNRs-based bipolar spin diodes. An external magnetic field is used to magnetize one or both GNR leads. *ML* and*MR* represent the magnetization of the left and right leads under the magnetic field, respectively. The value of*ML* and*MR* can be 1, 0 or −1, corresponding to magnetization along +*y* direction, non-magnetization, and magnetization along −*y* direction, respectively. (a) Under a positive bias, only spin down electrons transport through devices. The flow direction of electrons is from right to left leads. (b) Under a negative bias, only spin up electrons are allowed to transport from left to right leads. It behaves as a bias-controlled bipolar spin diode device. The circuit diagram of this bias-controlled bipolar spin diode is shown in the inset. (c) *I*−*V* curves of [1;1] and [1;−1] magnetic configurations of 8-ZGNR. The bias dependent spin polarization in the [1;−1] configuration are shown in inset. (d) and (e) *I* −*V* curves for 8-ZGNR with the magnetic configuration of [1;0] and [−1;0]. The inset is the enlarged *I*−*V* curves under a small bias. (f) Band structure for the left lead (left panel), transmission curve (middle panel), and band structure for the right lead (right panel) for spin up (blue curve) and spin down (orange curve) states at zero bias. The dashed (solid) lines with arrow, used to guide eyes, indicate the forbidden (allowed) hopping for electrons from the left lead to the right lead due to the symmetry mismatching (matching) of π and π∗ subbands. (g) The same configuration as (f) except in the presence of a positive bias (+0.4 V). Transmission gap↓(gap↑) is reduced (increased) compared with that in (f), which makes spin down channel open and spin up channel suppressed. (h) and (i) Isosurface plots of the G-point wave functions of π∗ and π subbands for 8-ZGNR. Red and Blue indicate opposite signs of the wave function.

Following the conventional approach used in GMR spintronic devices, we consider first the spin polarized current in the two-terminal device under the magnetic configurations [$M_L;M_R$] = [1;1] and [1;−1], respectively. In the [1;1] configuration, typical metallic behavior was found for both positive and negative bias. Under the positive bias, the transport and electronic properties of the system are in agreement with those of Ref. [11]. Interestingly, in the [1;−1] configuration, the spin transport property under a negative bias is completely opposite of that under a positive bias and the device functions as a bipolar spin diode. Under a positive bias, only the spin-down current is allowed to pass through the device, while under a negative bias, only the spin-up current is possible [see Fig. 1c]. The device is referred as a bipolar spin diode because its operation involves both spin-up and spin-down polarized currents which is different from the traditional bipolar devices based on electron and hole carrier types or polarized and unpolarized states in graphene-based memory devices.[24]

Besides the traditional parallel ([1;1]) and antiparallel ([1; −1]) magnetic configurations of the electrodes, we also considered spin dependent transport in the magnetic configurations [1;0] and [−1;0], *i.e.,* one electrode is magnetized by an applied magnetic field while the other is non-magnetic, which has not be considered in previous studies. Compared to the [+1,-1] magnetic configurations, the biggest advantage of the [±1; 0] configurations is that it enables reduction of design complexity since only one electrode needs to be magnetized. Moreover, it eliminates direct interaction of two magnetic electrodes which may be significant enough to affect the performance of nano-devices. The calculated *I-V* curves shown in Fig. 1d and 1e clearly indicate a bipolar spin diode behavior in these configurations. It is similar to that in the [1;−1] configuration [Fig. 1c], with the only difference being the zero threshold voltage in the [1;0] and [−1;0] configurations [see the inset of Fig. 1d]. This unique *I-V* characteristics suggests that the spin-up or spin-down conducting channel can be switched by either the bias direction or the magnetization of the ZGNR-electrode, as schematically illustrated in Fig. 1a and 1b. It can be explained based on the calculated electronic structure and symmetry of the electron wave function. Figures 1f and 1g show the calculated band structures of the left and right electrodes for the [1;0] configuration. The $\pi$ and $\pi*$ subbands in the right electrode are spin degenerate and cross the Fermi level without magnetization. The shift of energy bands by a small positive bias will open (close) the spin down (up) channel due to the symmetry matching (mismatching) of spin down (up) subbands in the left and right electrodes. The symmetry mismatching of $\pi*$ and $\pi$ subbands is reflected by their wave functions which is shown in Fig. 1h and 1i. As can be seen, the $\pi*$ subbands have the $\sigma$ symmetry, while the $\pi$ subbands show $C_2$ symmetry.

## Spin transistor

The above results of spin diodes indicate that the magnetic configurations, [1;0] and [−1;0], of the two-terminal device exhibit spin diode behavior and the conducting spin channel can be selected by setting proper bias direction (+ or −) and/or magnetic configuration (1,0 and −1) of the electrodes. This flexible control over spin current makes the two-terminal device a possible basic component for building multi-functional spintronic devices, such as spin transistors i.e. current/voltage amplifiers. The first device we propose here is a current amplifier. Two designs of the transistor, one for spin-up current and one for spin-down current, are shown in Fig. 2a and 2b, respectively, and a side view of the device is shown in Fig. 2c. Each transistor consists of three terminals, an emitter, a collector and a base. The only difference between the two designs is the magnetic configurations of the electrodes which are indicated in the figure. The emitter is grounded and the collector voltage ($V_C$) is fixed to a positive value, while the voltage of the base ($V_B$) can be varied which controls the flow and polarization of the current. The polarity of spin transistor is determined by the magnetic configuration between the emitter and the collector. In Fig. 2a, the magnetic configuration between the emitter and the base is [1;0] and the bias voltage is negative. Based on the *I-V* property shown in Fig. 1d, the current $I_E$ contains only spin-up electrons. Between the collector and the base, the magnetic configuration is [−1;0] and the bias voltage is positive as long as $V_B < V_C$, according to the *I-V* curve shown in Fig. 1e, only spin-up electrons are allowed to flow from the base to the collector. Similarly, the design shown in Fig. 2b allows only spin-down electrons to pass through the device. The equivalent circuit diagrams are shown in Fig. 2d and 2e, respectively. According to the *I-V* curve shown in Fig. 1d and 1e, $I = GV$ under a small bias (-0.4~0.4 V), where $G$ is the conductance. The current gain is defined as

$$\left|\frac{I_C}{I_B}\right| = \left|\frac{I_C}{I_E - I_C}\right| = \left|\frac{G(V_C - V_B)}{G(V_B - V_E) - G(V_C - V_B)}\right| = \left|\frac{1 - V_B/V_C}{2V_B/V_C - 1}\right| \tag{1}$$

The current gain ($I_C = I_B$) based on the above equation is shown as a function of $V_B = V_C$ in Fig. 2f. The current gain is equal to 1 when $V_B = 0$ and goes to zero when $V_B = V_C$, but it increases dramatically near $V_B = V_C = 1/2$, which suggests a high-performance ZGNR-based three terminal spin transistor. The amplification of spin-polarized current is also useful for detection of spin-polarized current.

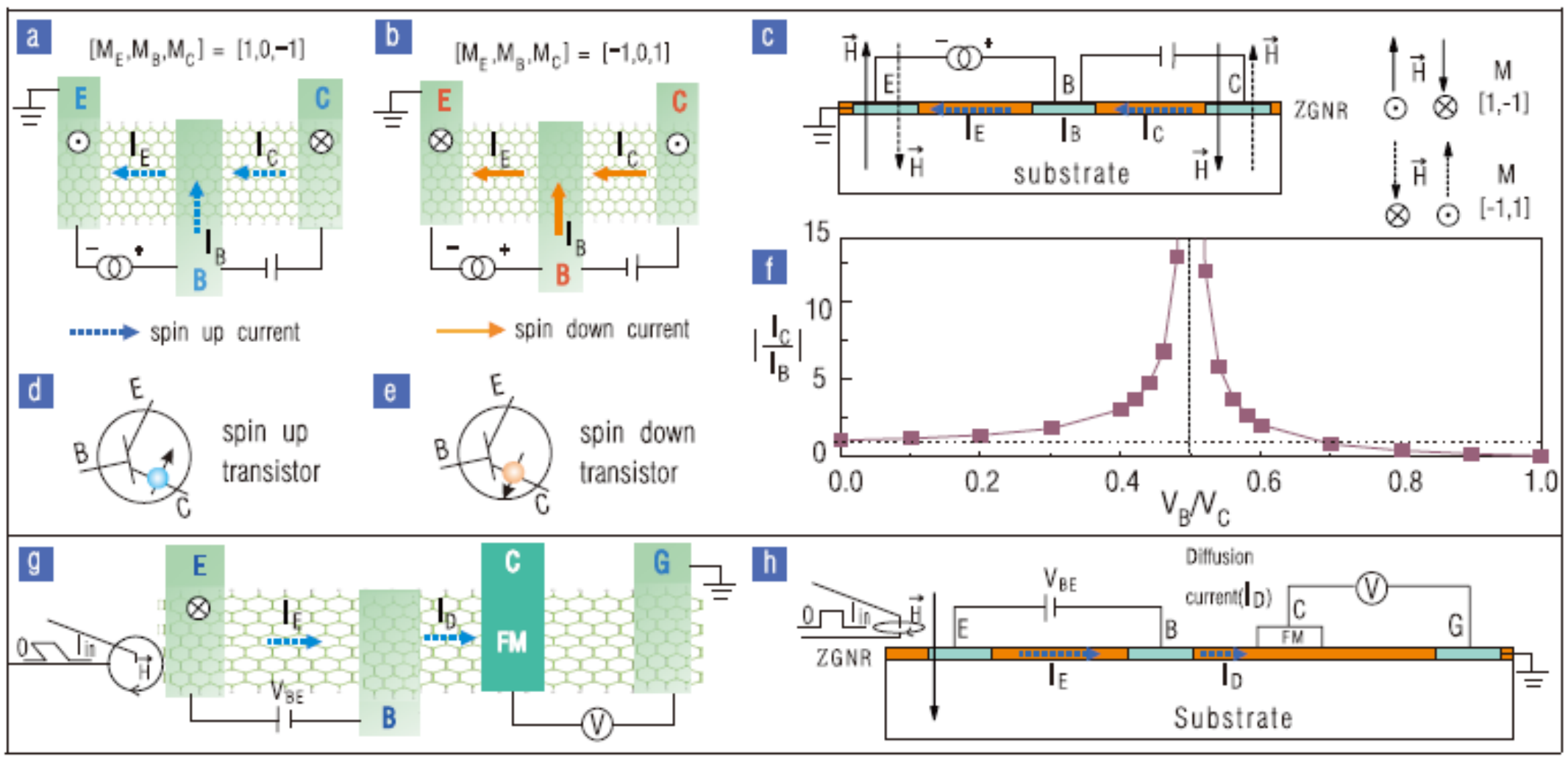

Figure 2: (Color online) top- and side-view of two types of transistors with the current and voltage amplification function. (a)-(c) Top- and side-view of ZGNR-based three-terminal spin up (down) transistors. (d) and (e) The circuit symbols of spin up (down) transistors. (f) The current gain (|$IC=IB$|) as function of$VB=VC$. (g) and (h) Top- and side-view of schematic illustrations of Johnson-type transistors as a voltage amplifier.

Transistors operating as voltage amplifiers can be designed similarly. In Fig. 2g and 2h, we show a possible Johnson-type of design.[25] Here the emitter and base are similar to those in the current amplifiers, but the collector is replaced by a cobalt and graphene junction. Note that a ferromagnetic cobalt electrode is used here to detect the spin-polarized current (up or down). The spin polarized current will flow from the emitter to the base under a bias voltage, and then diffuses from the base to the collector ($I_d$), generating a voltage difference ($V_d$) across the cobalt and graphene junction which can be measured. At a fixed diffuse distance, $V_d$ is determined by the injected spin polarized current at the base, which is linearly dependant on $V_{BE}$. Therefore we can have $V_d = \alpha V_{BE}$, where a is a constant. On the other hand, the magnetization of the emitter can be manipulated by placing a conducting wire with a current $I_{in}$ next to it. Thus the voltage gain of this transistor is

$$b = V_d = V_{in} = (\alpha V_{BE}) = (I_{in} R_{in}).$$

With proper choices of the parameters, it is possible for the transistor to act as a voltage amplifier. Compared with the conventional spin transistors, ZGNR-based spin transistors possess several unique features. First, the injected current from the emitter to the base is completely spin polarized. Second, graphene is a much better spin transport medium because the spin diffusion length in graphene is much longer than that in metal. In addition, the spin polarization of the current can be simply tuned by switching either the magnetization of the emitter-ZGNR or the DC bias direction, providing adjustable bipolar spin transistors.

## Spin logic gate

As discussed above, the flow of spin polarized current in a ZGNR-based device can be controlled by the magnetic configuration of the electrodes. This property of ZGNR can be used to construct a complete set of spin logic gates. In the following, we label the input terminals of the devices by *A* and/or *B*, and the output terminal by *Y*. In all designs, the logic inputs are encoded by the magnetization of the terminals, with positive magnetization of the ZGNR electrode representing the logic input 1 and negative magnetization representing logic 0. The result of the logic operation, however, is expressed in terms of the output current. For convenience of discussion, we assume that only the spin-up current is detected by non-local measurements,[4,25,26] so that we can encode the logic output to be 1 (0) if the output current include (exclude) the spin up current. Figure 3(a) shows schematic of the design for the NOT logic gate. The device consists of two terminals, and the magnetization of the right ZNGR electrode is fixed to zero (non-magnetic). The voltage of the left electrode is higher than that of the right electrode, so that the spin polarized current flows from left to right. If the magnetization of the left electrode is set to −1 (logic input 0), the spin-up channel is conducting, corresponding to a logic output 1. On other hand, the spin-down channel is conducting (logic output 0) when the magnetization of the left electrode is set to 1 (logic input 1). The NOT logic operation is thus realized. The truth table and circuit symbol are shown in Fig. 3a. Similarly, an AND gate can be designed but it requires three terminals, as shown in Fig. 3b. The magnetization of the left electrode is pinned to 1, and inputs are represented by the magnetization of the center and right terminals. The electric potential decreases from left to right. Based on the *I*-*V* curve given in Fig. 1, it can be easily seen that only when the magnetization of both the center and right terminal correspond to logic 1, the output includes the spin up current (logic output 1). In all other cases, either the spin polarized currents are completely blocked or only the spin-down current reaches the output terminal, corresponding to logic output 0. The logic operations are summarized in Fig. 3b, along with the truth table and circuit symbol of this AND gate. The logic OR operation can also be realized similarly as shown in Fig. 3c. Here, the left and right electrodes are used as the input terminals, and the middle electrode is used as the output terminal and its magnetization is pinned to 0. The spin-up current passes through the output terminal when magnetization of either or both input terminals is 1. Only when both of the input terminals are set to −1 (logic input 0), the output current consists of spin-down electrons, corresponding to logic output 0. Here, we only show the three base logic gates, NOT, AND, and OR. Actually, other logic operations can also be realized based on the design concepts above [see Support Information Figure 2] or the combination of these base logic gates. For example, the NAND logic gate can be realized by combining an AND gate and a NOT gate, the XOR logic gate can be realized by combining a NAND gate with a OR gate.

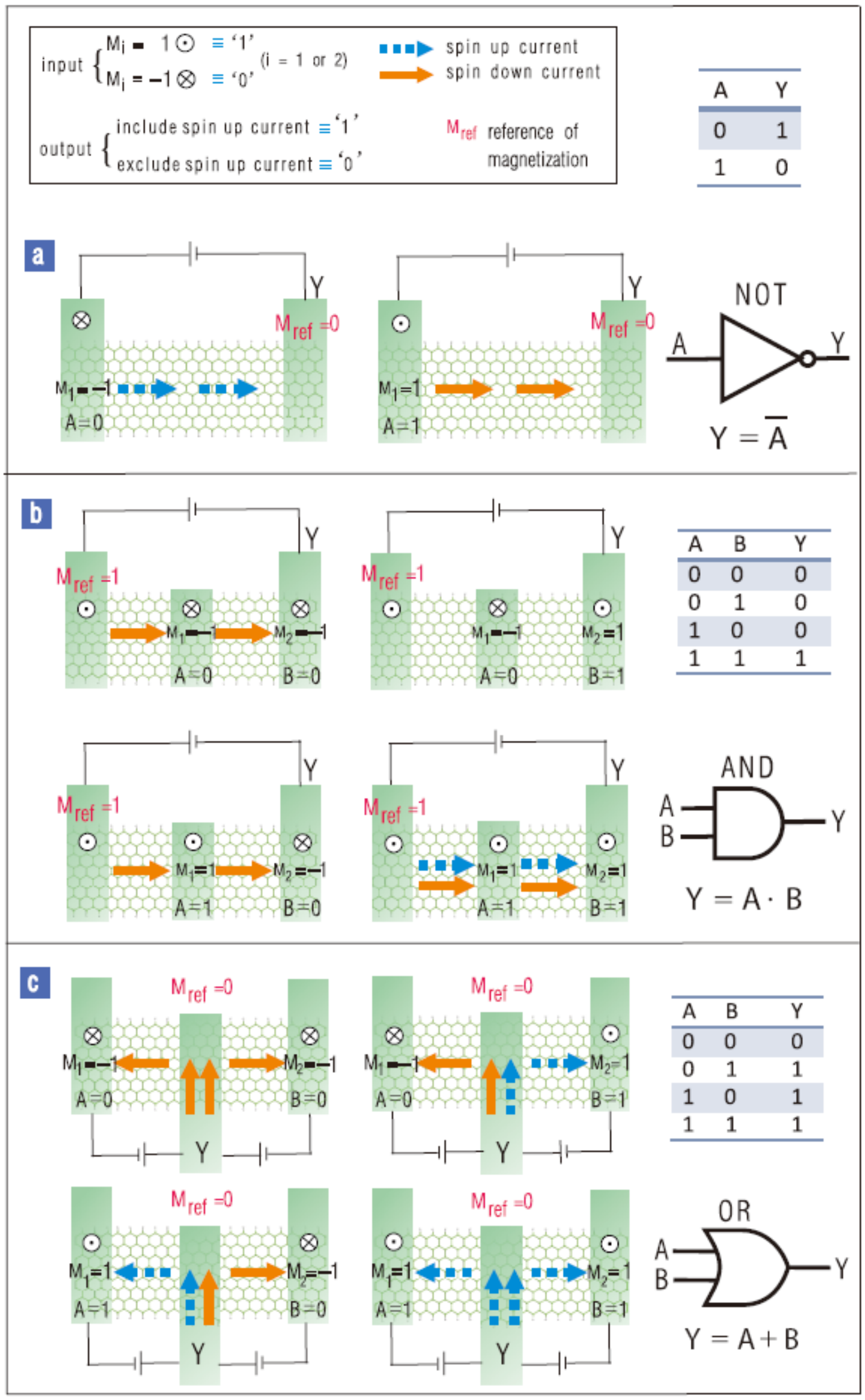

Figure 3: (Color online) Schematic illustrations of the spin logic gates: (a) Logic NOT gate (b) Logic AND gate (c) Logic OR gate. The input terminals are labeled by A and B; the output terminal is labeled by Y. *Mre f* represent the pinned magnetization of the terminal. The logic input 1 (0) is encoded by the magnetization 1 (-1) of the input terminals. The logic output 1 (0) is encoded if the output current include (exclude) the spin up current. The truth table and circuit symbol are shown in the right side of each panel.

## Spin calculator

As a potential device for molecular scale computation, molecular scale calculators (moleculators) have attracted much interest of chemists.[27] Graphene may be the best material for moleculators due to its thermal stability and weaker spin-orbital and hyperfine interactions. To realize graphene based moleculators, the design of graphene-based logic gate that can perform fundamental Boolean functions is essential. Here, we present a possible design of a moleculators based on the ZGNR-based spintronic logic gates discussed above. We mainly demonstrate a half adder here since the full adder is a simple combination of two half adders, and subtraction is simply an addition of negative numbers, the multiplication is repeated addition and division is repeated subtraction. A half adder is a logical circuit that performs an addition operation on two one-bit binary numbers often written as A and B. The half adder output is a sum of the two inputs usually represented with the signals C and S where $sum = 2 \times C + S$. Figure 4 shows the schematic diagram, the circuit symbol and the truth table for a half adder. As can be seen, the half adder is composed of a XOR and AND logic gates. Since graphene-based devices can be cut out from a piece of whole graphene sheet,[28] it eliminates the interconnection problem of two logic gates as shown in Fig.4a. Note that in the logic device design, both the input and output signals of logic gates can be used by current due to the electromagnetic conversion, such as using a solenoid to generate magnetization.

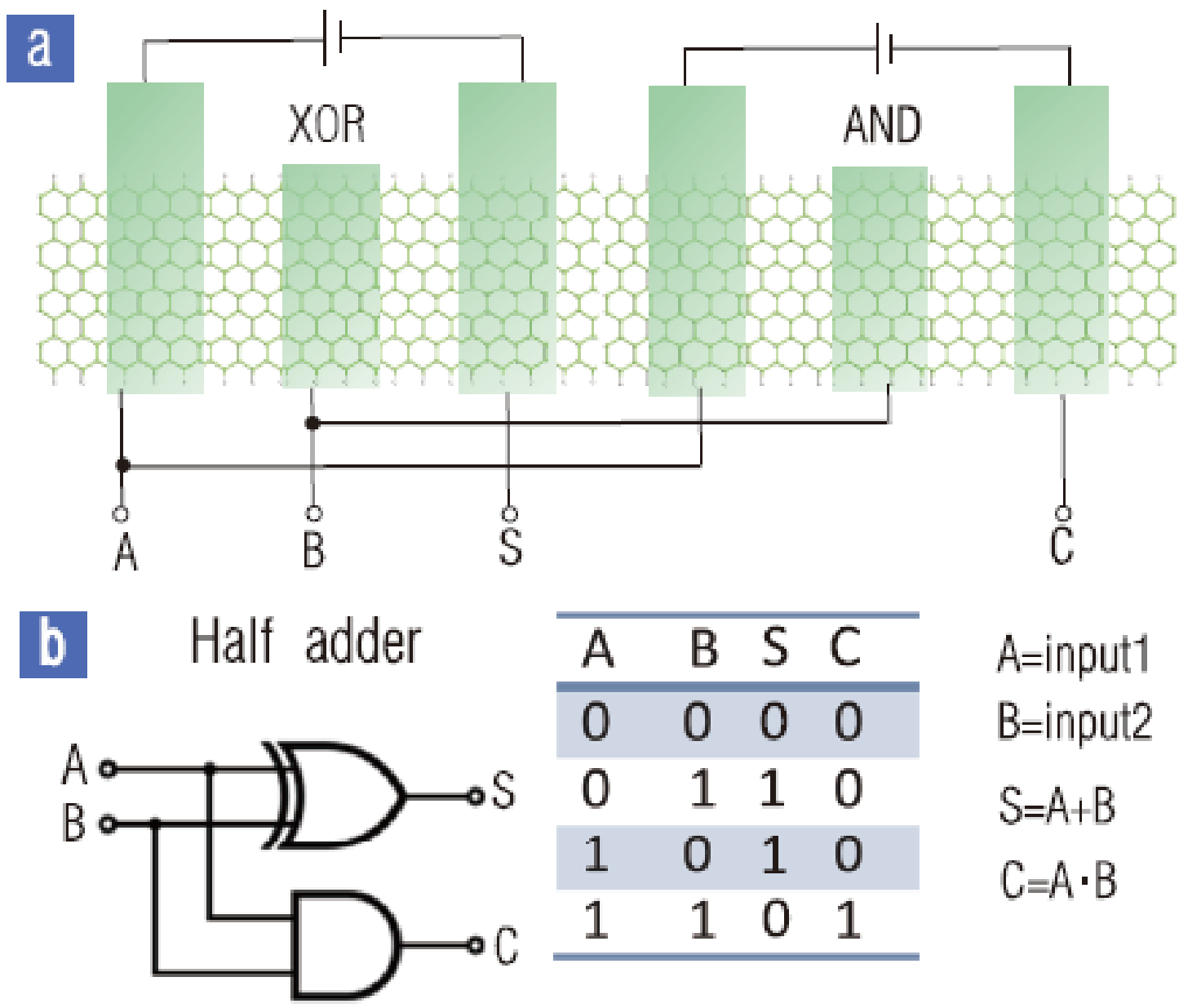

Figure 4: (Color online) Schematic diagram of a half-adder and its true table.

## Conclusion

In summary, our first-principles studies on spin-dependent transport properties show that a bias-controlled bipolar spin diode behavior is an intrinsic property of biased ZGNRs. Spin polarization and flow direction of current can be controlled through the source-drain voltage and magnetic configuration of the electrodes. The freedom in controlling the spin-polarized current in ZGNR-based spin diodes allows us to theoretically design spin transistors and logic gates. These spin components can perform basic functions such as rectification, amplification and logic operation and make it possible to manipulate spin-polarized current in ZGNR-based spintronic devices and then the design of spintronics logic devices.

## Acknowledgement

The authors thank Michael B. Sullivan for his helpful discussion. Minggang Zeng thanks NanoCore at the National University of Singapore for providing Research Scholarship. This work is partially supported by the National Research Foundation (Singapore) Competitive Research Program (Grant No. NRF-G-CRP 2007-05) and A*STAR SERC grant (Grant No. M47070020).

## Supporting Information Available

Computational details; width dependent transmission spectrum and I-V curve; and NAND and NOR logic gates. This material is available free of charge via the Internet at http://pubs.acs.org/.